# Fractal Collapse in Developed Hydrodynamic Turbulence

Vladimir M. Malkin

*Institute for Advanced Study, Princeton, NJ 08540 & Budker Institute of Nuclear Physics, Russia*

The problem of intermittency in developed hydrodynamic turbulence is considered. Explicit formulae taking into account effects of finite size of the inertial range are presented for the whole set of intermittency exponents. The formulae fit pretty well experimental data whose apparent discrepancies are attributed to different sizes of the inertial ranges in different experiments. Further predictions are given that can be verified by already existing experimental technique.

• **INTRODUCTION:** Developed hydrodynamic turbulence was studied by many scholars of our century. The subject is complicated and a theory that fits a broad experimental database, having clear physical background and real predictive power is still wanted. The problem is known to be closely connected with the problem of blow-up of velocity gradients in the Euler fluid. The latter problem is also complicated and unsolved yet. Nevertheless, the joint study of these problems can be helpful and based on it one can arrive to a new qualitative picture of the intermittency and also to new quantitative results.

• **INTERMITTENCY EXPONENTS:** Let $\epsilon(\vec{r})$ be the local density of energy dissipation rate in a strongly turbulent fluid. The time argument $t$ is omitted here and below as all quantities considered at the same time $t$ and as the forthcoming space averaging removes all time dependence. The energy dissipation rate can be smoothed over a spatial scale $x$ by means of a "filter function" $F(\vec{\rho})$: $\epsilon(\vec{r}, x) = \int d\vec{\rho} F(\vec{\rho}) \epsilon(\vec{r} + x\vec{\rho})$, $\int d\vec{\rho} F(\vec{\rho}) = 1$. The result depends on the filter $F$. This dependence is important and will be described elsewhere. For now, $x$-dependence is considered assuming that the filter satisfies some basic requirements. For $x$ inside the inertial range, i.e., for $L \gg x \gg h$, where $L$ is the energy-containing and $h$ is dissipative (Kolmogorov) scale of the turbulence, some scale-invariance properties are expected. In particular, it is assumed that for $x$ and $y$ both inside the inertial range

$$\langle (\epsilon(\vec{r}, y)/\epsilon(\vec{r}, x))^q \rangle = (x/y)^{\mu_q(x/y)} , \qquad (1)$$

where the angle brackets signify space averaging. The scale invariance hypothesis is that the exponents $\mu_q(x/y)$ depend just on the ratio $x/y$ rather than on both variables $x$ and $y$. These exponents are usually referred to as the intermittency exponents, because they all are zero in the Kolmogorov model of turbulence [1] that assumes spatially uniform energy dissipation $\epsilon$. There are several definitions of the intermittency exponents in the literature, which sometimes assumed to be equivalent, but, strictly speaking, should be distinguished from each other.

It is convenient for further applications to change variable so that $x/y = 2^n$ and $\mu_q(x/y) = \mu_{n,q}$. As $n \to \infty$ (which limit makes sense just for infinitely broad inertial range), the intermittency exponents become scale-independent: $\mu_{n,q} \to \mu_q$. There are many experiment-based evaluations of $\mu_q$ with widely scattered results (and there are many phenomenological models trying to fit these evaluations). This is basically due to effects of finite size of the inertial range, which are important in all realistic situations. The importance of finite size corrections to the intermittency exponents strongly complicates the problem. It is possible however to introduce physical quantities which are less sensitive to the finite size effects than intermittency exponents.

• **FRACTAL DIMENSIONS:** Consider the spatial domain $H(\alpha, x, y)$ defined by condition

$$\epsilon(\vec{r}, y)/\epsilon(\vec{r}, x) \geq (x/y)^\alpha . \qquad (2)$$

Its average volume per cube $x^3$ depends on $y < x$ as

$$V(\alpha, x, y) = y^3 (x/y)^{D(\alpha, x/y)} = x^3 (x/y)^{D(\alpha, x/y)-3} , \quad (3)$$

which employs the scale invariance hypothesis again. The quantity $V/y^3$ can be interpreted as the smallest average number of cubes $y^3$ per cube $x^3$ required to cover domain $H(\alpha, x, y)$. Then, $D(\alpha, x/y) \equiv D_n(\alpha)$ is the dimensionality of $H(\alpha, x, y)$. It cannot exceed 3 and it is a decreasing function of $\alpha$. There is $\alpha_{n,0}$, such that $D_n(\alpha) = 3$ for $\alpha \leq \alpha_{n,0}$ and $D_n(\alpha) < 3$ for $\alpha > \alpha_{n,0}$. There is also $\alpha_{n,M}$, such that $D_n(\alpha) > 0$ for $\alpha < \alpha_{n,M}$ and $D_n(\alpha) \leq 0$ for $\alpha \geq \alpha_{n,M}$. It follows from (1)-(3) that, for a positive $q$,

$$\mu_{n,q} \geq \alpha q + D_n(\alpha) - 3 . \qquad (4)$$

The difference between left-hand side of (4) and its right-hand side is a non-negative function of $q$ (and other parameters). It has a minimum over $q$. Provided $\alpha$ is such that the minimum is reached at a finite positive $q$, its position is determined by $\mu'_{n,q} = \alpha$ (the prime signifies $q$-derivative) and its value is zero, so that there

$$D_n(\alpha) = 3 + \mu_{n,q} - q\mu'_{n,q} \equiv D_{n,q} . \qquad (5)$$

At $n \to \infty$ the dimensionality $D_{n,q}$ tends to its scale-independent limit $D_q$. There are other definitions of fractal dimensions $D_q$ in the literature which should be distinguished from that given above.

Formula (5) expresses $D_{n,q}$ in the terms of $\mu_{n,q}$. When $D_{n,q}$ is known, $\mu_{n,q}$ can be restored by

$$\mu_{n,q} = q \int_0^q dq_1 (3 - D_{n,q_1})/q_1^2 + \alpha_{n,0}\, q\,, \qquad (6)$$

that takes into account condition $\mu_{n,0} = 0$. The integral in (6) converges at $q_1 \to 0$, as $D_{n,0} = 3$ (see (5) at q=0) and it is the maximum value of $D_{n,q}$. It also makes clear that the above $\alpha_{n,0}$ corresponds to $q = 0$. Another useful form of (6) is

$$\mu_{n,q} = q \int_1^q dq_1 (3 - D_{n,q_1})/q_1^2 + \mu_{n,1}\, q\,. \qquad (7)$$

The first intermittency exponent is usually assumed to be zero, but in the framework of given definitions there is no reason for conclusion $\mu_{n,1} = 0$ at all $n$, although it is correct at $n = \infty$.

• **HOT SPOTS:** First consider the limit of infinitely broad inertial range. As $n \to \infty$ the quantities $D_{n,q}$, $\mu_{n,q}$ and $\alpha_{n,q} \equiv \mu'_{n,q}$ tend to their extreme scale-independent values $D_q$, $\mu_q$ and $\alpha_q$ respectively. At fixed q, the quantity $N \equiv V/y^3$ (defined by (3) and interpreted after it) depends of $n$ like $N = 2^{n D_q}$.

There is a physical interpretation of such a dependence. Recall that $N$ is the number of sub-cubes with side $y = x/2^n$ per cube $x^3$, required to cover domain $H(\alpha, x, y)$ where (2) is satisfied. Condition (2) selects the sub-cubes in which the energy dissipation rate exceeds certain level, i.e. the "hot spots" of certain intensity. One can assume that, as $n$ increases, the hot spots split into those of smaller size which then do the same at larger $n$ producing finally a fractal structure. This picture agrees qualitatively with the clustering of energy dissipation in turbulence. There is also a natural dynamic process underlying such a picture. It is a branching blow-up process — fractal collapse of velocity gradients in Euler fluid. If branching happens on average after each increase $n$ by some $\Delta$, then it should be $N = 2^{n/\Delta}$ branches (i.e. hot spots) at given $n$. The comparison with the previous expression for the number $N$ shows, that $D_q = 1/\Delta$ can be interpreted as the branching coefficient for fractal collapse.

Convexity of $\mu_{n,q}$ as the function of $q$ implies that $\alpha_{n,q} \equiv \mu'_{n,q}$ is an increasing function of $q$. Since $D_n(\alpha)$ is a decreasing function of $\alpha$ and $D_n(\alpha_{n,q}) \equiv D_{n,q}$, one concludes that $D_{n,q}$ is a decreasing function of $q$. The physical interpretation is that the branching delays collapse and reduces the strength of singularities. Taking higher moments $q$ selects more singular, i.e. less branched structures. At sufficiently large $q$ (and $n \to \infty$), just perfect non-branched structures are selected that correspond to the fastest possible blow-up of velocity gradients in Euler fluid. For larger $q$ ($q \geq q_M$), one has $D_q = 0$, $\mu_q = \alpha_M q + 3$. The largest singularity exponent $\alpha_M$ can be determined from a nonlinear eigenvalue problem for self-similar blow-up solutions of Euler equations. This is not done yet because of severe technical difficulties. Moreover, even the dimensionality of singularities is under discussion. The above picture assumes that singularities are point ones. There are also phenomenological models of turbulence (cited below) which assume line or plane singularities.

• **QUANTITATIVE PREDICTIONS:** Provided certain universality properties of collapse branching, selection of more singular structures by taking higher moments $q$ should be a uniform process in $q$. Specifically, increasing $q$ by some value should cause a $q$-independent decrease in $D_q$, i.e., it should be $D'_q = -\gamma = const$, if the dimensionality is determined just by collapse branching. There is however a background that goes down to compensate for the increase of $\epsilon$ in hot spots and secure the energy flux conservation. Indeed, it is easy to see that $\alpha_0 = \mu'_0 < 0$. The background contribution to the dimensionality is important at small $q$. The dimensionality $q$-dependence at small $q$ is quadratic in $q$, since $q = 0$ is the point of $D_q$ maximum ($D_0 = 3$). This quadratic decrease in $q$ is turned into the linear one at higher $q$ where behavior of $D_q$ is already determined by the collapse branching. A simple formula for $D_q$ having such properties is

$$D_q = 3 - \gamma q^2/\sqrt{q^2 + a^2}\,. \qquad (8)$$

This formula is valid for $q \leq q_M$, while $D_q = 0$ for $q \geq q_M$,

$$q_M = 3/\gamma \left[1/2 + \left(1/4 + (a\gamma/3)^2\right)^{1/2}\right]^{1/2}\,. \qquad (9)$$

Substitution (8) in (7) at $n = \infty$ gives

$$\mu_q = \gamma\, q\, [\operatorname{arsh}(q/a) - \operatorname{arsh}(1/a)]\,, \qquad (10)$$

where $\operatorname{arsh}\xi = \ln(\xi + \sqrt{\xi^2 + 1})$ is the hyperbolic arcsin. There is a connection between parameters $a$ and $\gamma$ (which is specified below), so that all quantities at $n = \infty$ can be expressed in the terms of a single parameter, say $\gamma$.

Having $\mu_q$ known, one can proceed with the calculation of finite size corrections. Their calculation, taking into account real experimental conditions and filter functions will be considered elsewhere, while now the above heuristic arguments are to be extended to finite $n$. These lead to the formulae (8)-(10) again, but all the quantities become $n$-dependent and a non-zero $\mu_{n,1}$ appears in (10):

$$D_{n,q} = 3 - \gamma_n q^2/\sqrt{q^2 + a_n^2}\,, \qquad q \leq q_{n,M}\,, \qquad (11)$$

$$q_{n,M} = 3/\gamma_n \left[1/2 + \left(1/4 + (a_n \gamma_n/3)^2\right)^{1/2}\right]^{1/2}\,, \qquad (12)$$

$$\mu_{n,q} = \gamma_n q\, [\operatorname{arsh}(q/a_n) - \operatorname{arsh}(1/a_n)] + \mu_{n,1}\, q\,. \qquad (13)$$

These formulae contain three functions of one variable, $\gamma_n$, $a_n$ and $\mu_{n,1}$, unknown yet. The function $\mu_{n,1}$ is small and tends to zero at $n \to \infty$. The function $\gamma_n$ is constant up to a small varying term and given by

$$\gamma_n = \gamma + 2\mu_{n,1} \ . \qquad (14)$$

It is sufficient now to have formulae for $\mu_{n,1}$ and $\mu_{n,2}$ in order to determine all $\mu_{n,q}$. For derivation of scale dependence of intermittency exponents, quite different ideas are required which are considered in [6]. Here are formulae for $\mu_{n,1}$ and $\mu_{n,2}$ without derivation:

$$\mu_{n,1} = -\log_2 g_{n,1}/n \ , \quad g_{n,1} = 1 + \delta \left(1 - 2^{-bn}\right) ,$$
$$b = D_1 + 1 - \mu_2 - 2\delta \ , \qquad \delta \ll 1 ; \qquad (15)$$
$$\mu_{n,2} = \mu_2 - \log_2 g_{n,2}/n \ ,$$
$$g_{n,2} = g_2 + (1 - g_2)(G_2/2)^n + 2(g_{n,1} - 1) ;$$
$$g_2 = 1/(2 - G_2) \ , \quad G_2 = 2^{\mu_2} \ . \qquad (16)$$

The quantity $a_n$ is expressed explicitly in the terms of $\gamma_n$, $\mu_{n,1}$ and $\mu_{n,2}$ by

$$a_n = 2\left[A_n(2 - A_n)(2A_n - 1)\right]^{1/2}/\left(A_n^2 - 1\right) \ ,$$
$$A_n = \exp\left[(\mu_{n,2} - 2\mu_{n,1})/(2\gamma_n)\right] \ . \qquad (17)$$

The singularity exponents $\alpha_{n,q} \equiv \mu'_{n,q}$ at $n = 1; \infty$ and $q = 0; q_{n,M}$ satisfy the relations

$$2^{3-\alpha_{1,0}} = 2^{\alpha_{1,M}} + 7 \ , \qquad 2^{3-\alpha_0} = 2^{\alpha_M} + 7 - 2\delta \ . \qquad (18)$$

These formulae allow one to express all $\mu_{n,q}$ in the terms of the single parameter $\gamma$.

• **COMPARISON WITH EXPERIMENTS AND PHENOMENOLOGICAL MODELS:** The experimental intermittency exponents $\mu_{n,q}$ for $q$ ranged from 2 to 10 and the scale ratios $x/y$ from 1.5 to 32 (which corresponds to $n$ ranged from 0.6 to 5) are presented in paper [2]. The best fit to experimental $\mu_{n,2}$ is reached at $\gamma = 0.296$. The corresponding values of the most important other parameters, calculated from the above formulae, are: $\mu_2 = 0.214$; $\delta = 0.023$; $b = 3.623$; $\alpha_M = 0.817 \approx 4/5$; $\alpha_0 = -0.124 \approx -1/8$.

The theoretical intermittency exponents $\mu_{n,q}$ as functions of $x/y = 2^n$ at $q = 1, 2, ..., 10$ are presented in the Fig.1-top (the order is clear as $\mu_{n,q}$ is increasing function of $q$). The circles are the experimental $\mu_{n,q}$ from Fig.7 of [2].

The theoretical intermittency exponents $\mu_{n,q}$ as functions of $q$ for $x/y = 2^n = 1, 1.5, 2, 4, 8, 32, \infty$ are presented in the Fig.1-bottom by solid lines (the order is clear as $\mu_{n,q}$ is increasing function of $n$). The circles are the experimental $\mu_{n,q}$ from Fig.7 of [2]. The pluses are $\mu_q = q - \zeta_{3q}$, where $\zeta_p$ are the experiment-based velocity structure functions exponents from the Fig.3 of [3]. The 'x-marks' are experimental $\mu_q$ from [4] (see below concerning the negative $q$). The dash-dot line is the Meneveau–Sreenivasan (MS) phenomenological model [4] corresponding in current notations to $D_\infty = 2, \alpha_\infty = 1/2$ (parameter $\alpha_\infty$ is slightly adjusted, so that $\mu_2 = 0.228$, rather $\mu_2 = 0.25$ as in [4]). The dashed line is the She–Leveque (SL) phenomenological model [5] based on other set of experiments and corresponding in current notations to $D_\infty = 1, \alpha_\infty = 2/3$.

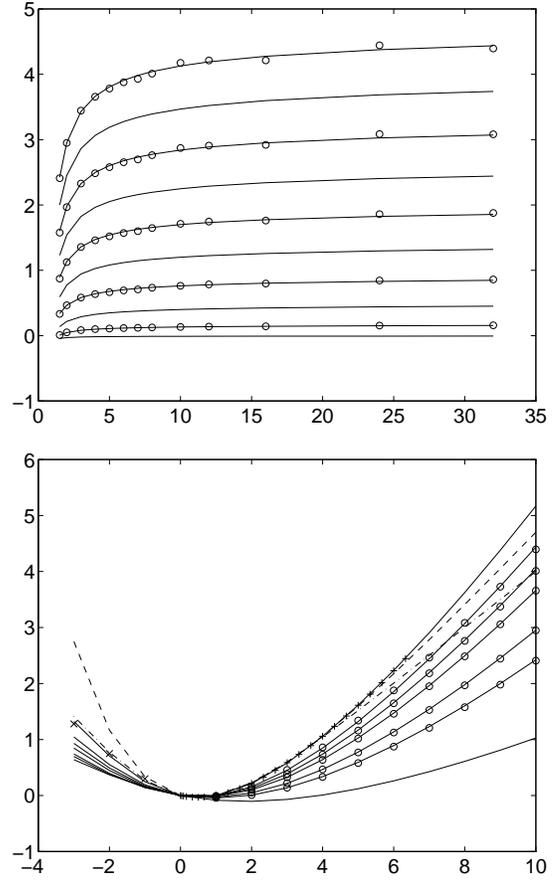

FIG. 1. Intermittency exponents $\mu_{n,q}$ as functions of $2^n$ – top, and of $q$ – bottom (see text for details).

As is clear from the Fig.1-bottom, the limit $n = \infty$ of the current theory is practically indistinguishable from both the SL and adjusted MS models in the range $0 \le q \le 4$. This range includes the whole range of data (approximately $0 \le q \le 3$) that was used to justify SL-model. Noteworthy, that the data of [2] differ substantially from the $x/y = \infty$-limit even at $x/y = 32$, where, for instance, $\mu_{n,2} \approx 0.16$. There is no contradiction however according to the current theory, as the finite size corrections $\mu_q - \mu_{n,q}$ decrease very slowly at $n \to \infty$.

The theoretical dimensionalities $D_{n,q}$ as functions of $q$ for $x/y = 2^n = 1, 1.5, 2, 4, 8, 32, \infty$ are presented in the Fig.2 by solid lines (the order is clear as $D_{n,q}$ is a decreasing function of $n$). The dash-dot and dashed lines represent the dimensionality $D_q$ defined as above and calculated according the formulae for $\mu_q$ of the MS and SL models respectively.

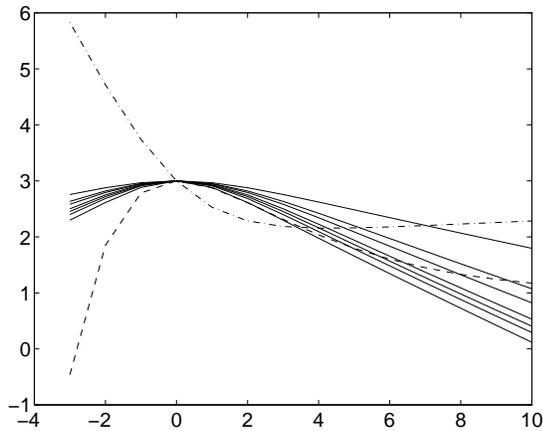

FIG. 2. Dimensionalities $D_{n,q}$ as functions of $q$

- **NEGATIVE MOMENTS:** The negative $q$ could be considered in a similar manner as positive $q$, if one assumes that domains of low dissipation are associated with the same singular structures as hot spots. Such an assumption is justified for sufficiently anisotropic collapse, for which formation of hot spots includes substantial angular redistribution of local energy flux $\epsilon$ near a singularity, and thus produces a "cold spot" right near a hot spot. Then the number of cold spots of a given intensity should be the same as the number of hot spots of the corresponding intensity. The selection of colder spots by taking higher negative moments $q$ should be a uniform process in $q$, similar to that considered above for positive $q$. As a result similar formulae arise. If one assumes that the first two derivatives of $D_{n,q}$ over $q^2$ are continuous at $q = 0$, it would imply that $D_{n,-q} = D_{n,q}$, i.e. $D_{n,q}$ is an even function of $q$.

Formula (6) for $\mu_{n,q}$ is applicable for $q < 0$ as well. For $D_{n,q}$ even in $q$, the first item in (6) is also an even function of $q$. Then the odd part of $\mu_{n,q}$ is a linear function of $q$, specifically, $\mu_{n,q} - \mu_{n,-q} = 2\alpha_{n,0} q$. Since $\alpha_{n,0} < 0$, the intermittency exponents for a negative $q$ are larger than for the positive $q$ of the same absolute value.

Thus extended to negative $q$ the dimensionality and intermittency exponents are presented in Fig.2 and Fig.1-bottom, as well as the extensions of MS and SL models. The SL model does not pretend to describe negative $q$ and it is not related to experimental data in this region. The MS model is desired to fit experimental $\mu_q$ for all $q$ including negative ones. As is seen from Fig.1-bottom, the theoretical $\mu_q$ agrees with the MS experimental data of [4] for negative $q$ as good as the MS model.

- **FEW REMARKS:** It is important to determine the experimental $\mu_{n,q}$ with a small step in $q$ (which is a problem of the data processing, rather than the measurements). This would allow one to make an accurate differentiation and find experimental $D_{n,q}$. The accuracy requirements are clear from the Fig.2 where the MS and my $D_q$ strongly diverge at negative $q$, although the MS and my $\mu_q$ are close to each other in the same region of Fig.1-bottom.

The first intermittency exponent $\mu_{n,1}$ is small, but it is important for the quantitative comparison between the theory and experiment. To the best of my knowledge, there is no direct experimental data on $\mu_{n,1}$. This is probably because the first intermittency exponent is usually assumed to be zero. It is desirable to determine the experimental $\mu_{n,1}$ and to compare it with the above predictions.

There is a fundamental problem of applying filters more general than one-dimensional cuts currently used in nearly all experiments.

When the intermittency exponents are known for all $q$, one can restore the probability distribution of the breakdown coefficients $M(\vec{r}, x, y) \equiv \epsilon(\vec{r}, y)/\epsilon(\vec{r}, x)$. It has a symmetry corresponding to the considered above hidden symmetry of the intermittency exponents.

There is a connection between the scale-dependent intermittency exponents and the velocity structure functions in developed turbulence.

- **ACKNOWLEDGMENTS:** Research supported in part by grants of Alfred P. Sloan Foundation and NEC Research Institute, Inc.. I am thankful to Profs. G. Pedrizzetti, E.A. Novikov and A.A. Praskovsky; W. de Water and J. Herweijer; K.R. Sreenivasan and other colleagues who sent me their experimental and numerical data, and to Prof. T. Spencer for questions and remarks which taking into account in the manuscript has simplified its understanding for broad circles of scholars.

[1] A.N. Kolmogorov, Dokl. Akad. Nauk SSSR, **30**(4), 301 (1941); A.M. Obukhov, Dokl. Akad. Nauk SSSR, **32**(1), 22 (1941); L.D. Landau and E.M. Lifshitz, *Fluid Mechanics*, translated from the Russian by J.B. Sykes and W.H. Reid. Oxford ; New York : Pergamon Press, 1993; A.S. Monin and A.M. Yaglom, *Statistical Fluid Mechanics*, Vol.2, MIT Press, Cambridge, MA, 1975.

[2] G. Pedrizzetti, E.A. Novikov and A.A. Praskovsky, Phys. Rev. E, **53**, 475 (1996).

[3] J. Herweijer and W. de Water, Phys. Rev. Lett. **74**(23), 4651 (1995).

[4] C. Meneveau and K.R. Sreenivasan, Phys. Rev. Lett. **59**(13), 1424 (1987); J.Fluid Mech., **224**, 429 (1991); K.R. Sreenivasan and G. Stolovitzky, J. of Stat. Phys., **78**(1/2), 311 (1995)

[5] Z.-S. She and E. Leveque, **72**, 336 (1994); B. Dubrulle, Phys. Rev. Lett. **73**, 959 (1994); Z.-S. She and E.C. Waymire, Phys. Rev. Lett. **74**, 262 (1995).

[6] V.M. Malkin, *"Scale dependence of intermittency exponents in developed hydrodynamic turbulence"* (to be published)